\documentclass[11pt]{article}
\usepackage{longtable,supertabular}
\usepackage{amsmath}
\usepackage{amssymb}
\usepackage{latexsym}
\usepackage{cite}

\title{\bf New Euclidean and Hermitian Self-Dual Cyclic Codes with Square-Root-Like Minimum Distances}
\author{Hao Chen
  \thanks{Hao Chen is with the College of Information Science and Technology/Cyber Security, Jinan University, Guangzhou, Guangdong Province, 510632, China, haochen@jnu.edu.cn. The research of Hao Chen was supported by NSFC Grant 62032009.}}

\begin{document}

\maketitle
\begin{abstract}
Binary self-dual codes with large minimum distances, such as the extended Hamming code and the Golay code, are fascinating objects in the coding theory. They are closely related to sporadic simple groups, lattices and invariant theory. A family of binary self-dual repeated-root cyclic codes with lengths $n_i$ and minimum distances $d_i \geq \frac{1}{2} \sqrt{n_i+2}$, $n_i$ goes to the infinity for $i=1,2, \ldots$, was constructed in a paper of IEEE Trans. Inf. Theory, 2009. In this paper, we construct families of Euclidean self-dual repeated-root cyclic codes over the field ${\bf F}_{2^s}$, $s \geq 2$,  with lengths $n_i$ and minimum distances at least $\sqrt{2^{s-1}n}-2^s$, where lengths $n_i$ go to the infinity.  We also construct families of Hermitian self-dual repeated-root cyclic codes over the field ${\bf F}_{2^{2s}}$, $s \geq 1$,  with lengths $n_i$ and minimum distances at least $\sqrt{n_i/2}$, where lengths $n_i$ go to the infinity. Our results show that Euclidean and Hermitian self-dual codes with large automorphism groups and large minimum distances can always be constructed.\\

{\bf Index terms:} Euclidean and Hermitian self-dual codes, Euclidean and Hermitian dual-containing codes, Cyclic code.
\end{abstract}

\section{Introduction and Preliminaries}

The Hamming weight $$wt({\bf a})=|\{i: a_i \neq 0\}|$$ of a vector ${\bf a}=(a_0, \ldots, a_{n-1}) \in {\bf F}_q^n$ is the number of non-zero coordinate positions. The Hamming distance $d({\bf a}, {\bf b})$ between two vectors ${\bf a}$ and ${\bf b}$ is $$d({\bf a}, {\bf b})=wt({\bf a}-{\bf b}).$$ For a code ${\bf C} \subset {\bf F}_q^n$, its Hamming distance is $$d({\bf C})=\min_{{\bf a} \neq {\bf b}} \{d_H({\bf a}, {\bf b}),  {\bf a} \in {\bf C}, {\bf b} \in {\bf C} \}.$$  Then it is clear that the minimum Hamming distance of a linear code is its minimum Hamming weight.  An $[n, k, d]_q$ code over ${\bf F}_q$ is a linear code with the length $n$, the dimension $k$ and the minimum distance $d$. The Singleton bound asserts that $d \leq n-k+1$ for a linear $[n, k, d]_q$ code. A linear code attaining this bound is called  maximal distance separable (MDS). Reed-Solomon codes are well-known MDS codes, see \cite{HP,Lint,MScode}. \\

The Euclidean inner product on ${\bf F}_q^n$ is defined by $$<{\bf x}, {\bf y}>=\Sigma_{i=1}^n x_iy_i,$$ where ${\bf x}=(x_1, \ldots, x_n)$ and ${\bf y}=(y_1, \ldots, y_n)$. The Euclidean dual of a linear code ${\bf C}\subset {\bf F}_q^n$ is $${\bf C}^{\perp}=\{{\bf c} \in {\bf F}_q^n: <{\bf c}, {\bf y}>=0, \forall {\bf y} \in {\bf C}\}.$$ A linear code ${\bf C} \subset {\bf F}_q^n$ is Euclidean self-orthogonal if ${\bf C} \subset {\bf C}^{\perp}$, is Euclidean  self dual if ${\bf C}={\bf C}^{\perp}$, is Euclidean dual-containing if ${\bf C}^{\perp} \subset {\bf C}$, and is Euclidean linear complementary dual (LCD) if ${\bf C}^{\perp} \cap {\bf C}={\bf 0}$. The Euclidean dual of a Euclidean dual-containing code is a Euclidean self-orthogonal code. Similarly the Hermitian inner product is defined on ${\bf F}_{q^2}^n$ by $$<{\bf x}, {\bf y}>_H=\Sigma_{i=1}^n x_iy_i^q,$$ where ${\bf x}=(x_1, \ldots, x_n)$ and ${\bf y}=(y_1, \ldots, y_n)$ are two vectors in ${\bf F}_{q^2}^n$. The Hermitian dual a linear code ${\bf C}\subset {\bf F}_{q^2}^n$ is $${\bf C}^{\perp_H}=\{{\bf c} \in {\bf F}_{q^2}^n: <{\bf c}, {\bf y}>_H=0, \forall {\bf y} \in {\bf C}\}.$$ A linear code ${\bf C} \subset {\bf F}_{q^2}^n$ is Hermitian self-orthogonal if ${\bf C} \subset {\bf C}^{\perp_H}$, is self dual if ${\bf C}={\bf C}^{\perp_H}$, is dual-containing if ${\bf C}^{\perp_H} \subset {\bf C}$, and is linear complementary dual (LCD) if ${\bf C}^{\perp_H} \cap {\bf C}={\bf 0}$. The Hermitian dual of a Hermitian dual-containing code is a Hermitian self-orthogonal code. For a linear code ${\bf C} \subset {\bf F}_{q^2}^n$, we set $${\bf C}^q=\{(c_0^q, \ldots, c_{n-1}^q): (c_0, \ldots, c_{d-1}) \in {\bf C}\}.$$ Then it is clear that $${\bf C}_{\perp_H}=({\bf C}^{\perp})^q=({\bf C}^q)^{\perp}.$$ Hence the Hermitian dual can be thought as the composition of the Euclidean dual and the $q$-th power operation.  For the construction of Euclidean self-dual MDS codes over ${\bf F}_{2^s}$, we refer to \cite{GG08,Chen}.\\

Binary self-dual codes with large minimum distances are among most fascinating objects in mathematics. They have been invented and studied in the early days of the coding theory. The Golay code found in 1949, see \cite{Golay}, was closely related to the Leech lattice found in 1967, see \cite{Leech}, and played a central role in the famous construction of Conway sporadic simple groups, see \cite{Conway1968}. For the theory of self-dual codes over small finite fields, we refer to \cite{CPS,CS,DGH,Rains}, \cite[Chapter 9]{HP} and \cite[Chapter 19]{MScode}.  It is always interesting in the coding theory to construct Euclidean and Hermitian self-dual codes with large minimum distances. It is well-known that there is a family of binary Euclidean self-dual codes meeting the Gilbert-Varshamov bound, and there is a family of Euclidean self-dual codes over ${\bf F}_q$ exceeding the Gilbert-Varshamov bound for $q \geq 64$ and $q \neq 125$, see \cite{Rains,Bassa}. \\

From the fundamental Gleason theorem, the minimum weight $d$ of a length $n$ binary self-dual code satisfies $$d\leq 4 \lfloor \frac{n}{24}\rfloor+6,$$ if $n \equiv 22$ $mod$ $24$, and $$d \leq 4 \lfloor \frac{n}{24}\rfloor+4,$$ otherwise. A binary self-dual code attains this upper bound is called an extremal self-dual code, see e.g. \cite[Chapter 19]{MScode}. The minimum weight $d$ of a length $n$ ternary self-dual code satisfies $$d\leq 3 \lfloor \frac{n}{12}\rfloor+3.$$ A ternary self-dual code attains this bound is called an extremal self-dual ternary code. It is well-known that extremal self-dual codes do not exist when lengths are large, see \cite{DGH}. The construction and the classification of extremal or optimal Euclidean or Hermitian self-dual codes over ${\bf F}_2$, ${\bf F}_3$ and ${\bf F}_4$ with small lengths $n \leq 136$, have been active, since the pioneering work of Conway and Sloane, we refer to \cite{DGH,Gulliver1,Harada,Gaborit,Gaborit1,Gaborit3,Dijk,Gulliver2,Bet}. On the opposite direction, it was proved in \cite{Carlet} that each linear code over ${\bf F}_q$, $q >3$, is equivalent to an Euclidean LCD code, and each linear code over ${\bf F}_{q^2}$, $q \geq 2$, is equivalent to a Hermitian LCD code. The hull-increasing variation problem of equivalent linear codes was proposed and studied in \cite{Chen}.\\

Let ${\bf C} \subset {\bf F}_q^n$ be a linear code. If $(c_0, c_1, \ldots, c_{n-1}) \in {\bf C}$, then $(c_{n-1}, c_0, \\ \ldots, c_{n-2}) \in {\bf C}$, this code ${\bf C} \subset {\bf F}_q^n$ is called cyclic. A codeword ${\bf c}$ in a cyclic code is identified with a polynomial ${\bf c}(x)=c_0+c_1x+\cdots+c_{n-1}x^{n-1}\in {\bf F}_q[x]/(x^n-1)$. Every cyclic code is a principal ideal in the ring ${\bf F}_q[x]/(x^n-1)$ and then generated by a factor of $x^n-1$. The code with the generator polynomial $g(x)=g_0+g_1x+\cdots+g_{n-k}x^{n-k} \in {\bf F}_q[x]$ is denoted by  ${\bf C}_g$. The dimension of the cyclic code ${\bf C}_g$ generated by $g(x)$ is $n-\deg(g)$. The dual code of a cyclic code ${\bf C}_g$ is a cyclic code with the generator polynomial $g^{\perp}=\frac{x^k h(x^{-1})}{h(0)}$, where $h(x)=\frac{x^n-1}{g(x)}$. Therefore the root of $g^{\perp}$ is of the form $\frac{1}{\beta}$ if $\beta$ is not a root of $g(x)$, where $\beta$ is a $n$-th root of $1$ in some extension field of ${\bf F}_q$, see \cite{HP} Chapter 4.\\

Duadic codes were introduced in 1981 by Leon, Pless and Sloane, see \cite{Leon}, as extensions of quadratic residue codes. It is well-known that duadic codes have the square root lower bound on their minimum distances, see \cite[Chapter 6]{HP}. In recent papers \cite{TangDing,SunDW,SunDing,Liu,ChenTing}, new binary cyclic codes, new cyclic or negacyclic ternary codes and new constacyclic codes with square-root-like  lower bounds on their minimum distances and dual minimum distances were constructed.\\

It was proved that extended (binary) quadratic residue codes (then an extended code of a cyclic code) with length $p+1$ and the minimum distance $\sqrt{p}$ are self-dual, where $p \equiv \pm 1$ $mod$ $8$, see \cite[Chapter 6]{HP}. Some extended duadic codes are self-dual codes with square-root minimum distances, see \cite[Chapter 6]{HP} and \cite{Leon}. In \cite{Kai,Jia}, it was proved that self-dual codes over ${\bf F}_q$ only exist for even $q$.\\

Binary self-dual cyclic codes were firstly studied by N. J. A. Sloane and J. G. Thompson in \cite{Sloane} in 1983. Self-dual cyclic codes are self-dual codes with large automorphism groups. Then it is interesting to construct Euclidean and Hermitian self-dual cyclic codes over small fields with large minimum distances. In 2009, there was an important progress that a family of binary self-dual cyclic codes with the length $n=2(2^{2a+1}-1)$ and the minimum distance $\delta \geq \frac{1}{2} \sqrt{n+2}$ were constricted, for $a=1,2 \ldots$, by B. Heijne and J. Top in  \cite{HTop}. To the best of our knowledge, there is no other construction of binary self-dual cyclic codes with a large lower bound on their minimum distances. In our previous paper \cite{Chen1}, a family of binary self-dual repeated-root cyclic codes with minimum distances at least $\sqrt{n}-2$ was constructed. Then the results in \cite{HTop,TangDing} were improved significantly.\\

In this paper, we first give the $[{\bf u}|{\bf u}+{\bf v}]$ construction of Euclidean and Hermitian self-dual codes from Euclidean and Hermitian dual-containing codes. Then families of Euclidean self-dual repeated-root cyclic codes over ${\bf F}_q$, $q=2^s$, $s \geq 2$, with the length $n=\frac{2(q^m-1)}{\mu}$, $\mu$ is a divisor of $q^m-1$,  and the minimum distance $\sqrt{\frac{q}{2\mu}}\sqrt{n}-q$ are constructed. In the case $q=4$, our quaternary Hermitian self-dual repeated-root cyclic codes with lengths $n=2(4^m-1)$, $m=3, 5, 7, \ldots$, and minimum distances at least  $\sqrt{n/2}$, contain the groups ${\bf Z}_{4^m-1} \times {\bf Z}_{4^m-1}$ in their automorphism groups. Therefore Euclidean and Hermitian self-dual codes with large automorphism groups and large minimum distances can always be constructed.\\

\section{The $[{\bf u}|{\bf u}+{\bf v}]$ construction of Euclidean and Hermitian self-dual codes}

We first give the main construction of self-dual codes over ${\bf F}_q$ from dual-containing codes.\\

{\bf Theorem 2.1.} {\em Let ${\bf F}_{2^s}$ be a finite field with $2^s$ elements. Suppose that ${\bf C} \subset {\bf F}_{2^s}^n$ is an Euclidean dual-containing code. Then the linear code $${\bf C}_1=\{{\bf u}|{\bf u}+{\bf v}: {\bf u} \in {\bf C}, {\bf v} \in {\bf C}^{\perp}\} \subset {\bf F}_{2^s}^{2n}$$ is a self-dual code with the minimum distance at least $\min\{d({\bf C}^{\perp}), 2d({\bf C})\}$. Let ${\bf F}_{2^{2s}}$ be a finite field with $2^{2s}$ elements. Suppose that ${\bf C} \subset {\bf F}_{2^{2s}}^n$ is an Hermitian dual-containing code. Then the linear code $${\bf C}_1=\{{\bf u}|{\bf u}+{\bf v}: {\bf u} \in {\bf C}, {\bf v} \in {\bf C}^{\perp_H}\} \subset {\bf F}_{2^{2s}}^{2n}$$ is an Hermitian self-dual code with the minimum distance at least \\$\min\{d({\bf C}^{\perp_H}), 2d({\bf C})\}$.}\\

{\bf Proof.} For two codewords ${\bf c}_1=[{\bf u}_1|{\bf u}_1+{\bf v}_1]$ and ${\bf c}_2=[{\bf u}_2|{\bf u}_2+{\bf v}_2]$ in this code ${\bf C}_1$, their Euclidean inner product is $$<{\bf c}_1,{\bf c}_2>=(1+1)<{\bf u}_1, {\bf u}_2>+<{\bf u}_1, {\bf v}_2>+<{\bf u}_2,{\bf v}_1>+<{\bf v}_1, {\bf v}_2>.$$ Since ${\bf C}^{\perp}$ is self-orthogonal, this inner product is zero. It is clear that the dimension of ${\bf C}_1$ is $$\dim({\bf C}_1)=\dim({\bf C}^{\perp})+\dim({\bf C})=n.$$ Then the first conclusion follows immediately. The second conclusion can be proved similarly.\\

Let us consider the case of $s=1$. Suppose that the length of the binary cyclic code ${\bf C}$ is odd and the generator polynomial of ${\bf C}$ is $g_1(x) \in {\bf F}_2[x]$. Since ${\bf C}^{\perp} \subset {\bf C}$, then the generator polynomial of ${\bf C}^{\perp}$ is $g_1(x)g_2(x) \in {\bf F}_2[x]$. From the classical result in \cite{Lint1}, the above construction is a repeated-root cyclic code with the generator polynomial $g_1^2(x) g_2(x)$ after a coordinate permutation. Therefore if ${\bf C}$ in Theorem 2.1 is a binary cyclic code with the odd length $n$, then the code ${\bf C}_1$ is a repeated-root binary self-dual cyclic code of the length $2n$. By checking the proof of Theorem 1 in \cite{Lint1}, we have the following extension immediately.\\

{\bf van Lint theorem (see \cite{Lint1}).} {\em Let $q$ be an even prime power and $n$ be an odd positive integer. Let ${\bf C}_1 \subset {\bf F}_q^n$ be a cyclic code generated by the polynomial $g_1(x) \in {\bf F}_q[x]$ and ${\bf C}_2 \subset {\bf F}_q^n$ be a cyclic code generated by $g_1(x)g_2(x) \in {\bf F}_q[x]$. Notice that $g_1$ and $g_2$ are divisors of $x^n+1 \in {\bf F}_2[x]$, which have no repeated root. Then the code ${\bf C}=\{({\bf u}|{\bf u}+{\bf v}): {\bf u} \in {\bf C}_1, {\bf v} \in {\bf C}_2\}$ is the repeated-root cyclic code of the length $2n$ generated by the polynomial $g_1(x)^2g_2(x)$ after a coordinate permutation.}\\

{\bf Proof.} The proof is the same as the proof of Theorem 1 in \cite{Lint1}. Let ${\bf a}=(a_0, \ldots, a_{n-1}) \in {\bf C}_1$ and ${\bf c}=(c_0, \ldots, c_{n-1}) \in {\bf C}_2$. Set ${\bf b}={\bf a}+{\bf c}=(b_0, \ldots, b_{n-1})$. Then ${\bf a}(x)=a_0+a_1x+\cdots+a_{n-1}x^{n-1}={\bf a}_{even}(x^2)+x{\bf a}_{odd}(x^2)$ can be divisible by $g_1(x)$. And ${\bf c}(x)=c_0+c_1x+\cdots+c_{n-1}x^{n-1}$ can be divisible by $g_1(x)g_2(x)$. Therefore ${\bf b}(x)=b_0+b_1x+\cdots+b_{n-1}x^{n-1}={\bf b}_{even}(x^2)+x{\bf b}_{odd}(x^2)$ can be divisible by $g_1(x)$.\\

Set $${\bf w}=(a_0, b_1, a_1, b_3, \ldots,b_{n-2}, a_{n-1}, b_0, a_1, \ldots, a_{n-2}, b_{n-1}).$$ Then
$${\bf w}(x)=[{\bf a}_{even}(x^2)+x^{n+1}{\bf a}_{odd}(x^2)]+[x{\bf b}_{odd}(x^2)+x^n{\bf b}_{even}(x^2)].$$ It is clear that $${\bf a}_{even}(x^2)+x^{n+1}{\bf a}_{odd}(x^2)={\bf a}(x)+x(x^n+1){\bf a}_{odd}(x^2)$$ can be divisible by $g_1(x)$, since $g_1(x)$ is a divisor of $x^n+1$. Since there is only even degree powers $x^{2h}$'s in ${\bf a}(x)+x(x^n+1){\bf a}_{odd}(x^2)$, this term can be divisible by $g_1(x)^2$, from the fact that $g_1(x)$ has no repeated root. The second term $x{\bf b}_{odd}(x^2)+x^n{\bf b}_{even}(x^2)={\bf b}(x)+(x^n+1){\bf b}_{odd}(x^2)$ can be divisible by $g_1(x)^2$, from a similar argument.\\

On the hand ${\bf w}(x)=(x^n+1){\bf a}(x)+{\bf c}(x)+(x^n+1){\bf c}_{even}(x^2)$ can be divisible by $g_2(x)$, then ${\bf w}(x)$ can be divisible by ${\bf g}_1(x)^2g_2(x)$. Then the code ${\bf C}$ is in the cyclic code of the length $2n$ generated by $g_1(x)^2g_2(x)$. Because both codes have the same dimension $2n-2\deg(g_1)-\deg(g_2)$. The conclusion follows immediately.\\

\section{Euclidean and Hermitian dual-containing BCH codes}

Set ${\bf Z}_n={\bf Z}/n{\bf Z}=\{0, 1, \ldots, n-1\}$. A subset $C_i$ of ${\bf Z}_n$ is called a cyclotomic coset if $$C_i=\{i, iq, \ldots, iq^{l-1}\},$$ where $i \in {\bf Z}_n$ is fixed and $l$ is the smallest positive integer such that $iq^l \equiv i$ $mod$ $n$. It is clear that cyclotomic cosets correspond to irreducible factors of $x^n-1$ in ${\bf F}_q[x]$. Therefore a generator polynomial of a cyclic code is the product of several irreducible factors of $x^n-1$. The defining set of a cyclic code generated by ${\bf g}$  is the the following set $${\bf T}_{{\bf g}}=\{i: {\bf g}(\beta^i)=0\}.$$ Then the defining set of a cyclic code is the disjoint union of several cyclotomic cosets. Set $${\bf T}^{-1}=\{-i: i \in {\bf T}\}, $$ and $${\bf T}^{-q}=\{-qi: i \in {\bf T}\}.$$ Then the defining set of the dual code is $({\bf T}^c)^{-1}$, where ${\bf T}^c={\bf Z}_n-{\bf T}$ is the complementary set. It is well-known that a cyclic code with the defining set ${\bf T}$ is Euclidean dual-containing if and only if ${\bf T} \cap {\bf T}^{-1}=\emptyset$, see \cite{LDL2}. Notice that for cyclic codes in ${\bf F}_{q^2}^n$, we have to use $q^2$-cyclotomic cosets. Then the defining set of a cyclic code in ${\bf F}_{q^2}^n$ is the disjoint union of several $q^2$-cyclotomic cosets. A cyclic code ${\bf C} \subset {\bf F}_{q^2}^n$ with the defining set ${\bf T}$ is Hermitian dual-containing if and only if ${\bf T} \cap {\bf T}^{-q}=\emptyset$.  BCH codes were introduced in 1959-1960, see \cite{BC1,BC2,Hoc}, for giving a lower bound on minimum distances of cyclic codes. From the BCH lower bound, we can construct Euclidean and Hermitian dual-containing BCH codes with square-root-like minimum distances.\\

The following lemma is useful in this paper.\\

{\bf Lemma 3.1 (see \cite{LDL2}).} {\em Let $q>1$ be a positive integer and $a,b$ be two positive integers, then $\gcd(q^a-1, q^b-1)=q^{\gcd(a,b)}-1$. $\gcd(q^a+1, q^b-1)=1$ if $q$ is even and $\frac{b}{\gcd(a,b)}$ is odd, $\gcd(q^a+1, q^b-1)=2$ if $q$ is odd and $\frac{b}{\gcd(a,b)}$ is odd, $\gcd(q^a+1, q^b-1)=q^{\gcd(a,b)}+1$ if $\frac{b}{\gcd(a,b)}$ is even.}\\

The following result follows from the BCH bound and the characteristic of dual-containing cyclic codes.\\

{\bf Theorem 3.1.} {\em Suppose that there is no integer $a$ in ${\bf Z}_n$ such that both $a$ and $-a$ in the same cyclotomic coset. Moreover there is no integer $a \in {\bf Z}_n$ satisfying that both $a$ and $-a$ is in the set ${\bf T}=C_1 \cup C_2 \cup \cdots \cup C_{\delta-1}$. Then we can construct a length $n$ Euclidean dual-containing cyclic code with the defining set ${\bf T}$, such that its minimum distance is at least $\delta$. Suppose that there is no integer $a$ in ${\bf Z}_n$ such that both $a$ and $-qa$ in the same $q^2$-cyclotomic coset. Moreover there is no integer $a \in {\bf Z}_n$ satisfying that both $a$ and $-qa$ is in the set ${\bf T}=C_1 \cup C_2 \cup \cdots \cup C_{\delta-1}$. Then we can construct a length $n$ Hermitian dual-containing cyclic code with the defining set ${\bf T}$, such that its minimum distance is at least $\delta$.}\\

{\bf Proof.} First of all from the condition that there is no integer $a \in {\bf Z}_n$ such that both $a$ and $-a$ are in the same cyclotomic coset, all cyclotomic cosets can be paired. Since there is no integer $a \in {\bf Z}_n$, such that $-a$ and $a$ are in the defining set ${\bf T}=C_1 \cup C_2 \cup \cdots \cup C_{\delta-1}$, the cyclic code with this defining set is a Euclidean dual-containing code.  The conclusion about Hermitian dual-containing cyclic codes can be proved similarly.\\

Let $\mu $ be a divisor of $q^m-1$, $m$ odd, and the length be $n=\frac{q^m-1}{\mu}$. If there are $a$ and $-a$ in the same cyclotomic coset, then $a(q^i+1) \equiv 0$ $mod$ $n$, where $i \leq m$. It is easy to verify that this is not possible from Lemma 3.1. We have the following two results.\\

{\bf Theorem 3.2.} {\em Let $n$ as above be the length. Then we can construct an Euclidean dual-containing BCH code ${\bf C}$ with the defining set ${\bf T}=C_1 \cup C_2 \cup \cdots \cup C_{\delta-1}$, where $\delta = \frac{q^{\frac{m+1}{2}}-q}{\mu}$. Then we have $$d({\bf C}) \geq \frac{q^{\frac{m+1}{2}}-q}{\mu}+1.$$}\\

{\bf Proof.} We only need to prove that there is no two positive integer $1 \leq u, v \leq  \frac{q^{\frac{m+1}{2}}-q}{\mu}$, satisfying $uq^t+v \equiv 0$ $mod$ $n$, where $t \leq m$. Without the loss of the generality, we can assume that $t \leq \frac{m-1}{2}$. Otherwise $t \geq \frac{m+1}{2}$, $uq^m+vq^{m-t}\equiv u+vq^{m-t} \equiv 0$ $mod$ $n$. Then $\mu(uq^t+v) \geq q^m-1$, the conclusion follows directly.\\

The above result can be found in \cite{Aly}. For Hermitian dual-containing BCH codes, we have the following result. The proof is similar.\\

Let $\mu $ be a divisor of $q^{2m}-1$, $m$ odd, and the length be $n=\frac{q^{2m}-1}{\mu}$. If there are $a$ and $-a$ in the same $q^2$-cyclotomic coset, then $a(q^{2i}+1) \equiv 0$ $mod$ $n$, where $i \leq m$. It is easy to verify that this is not possible from Lemma 3.1. We have the following result.\\

{\bf Theorem 3.3.} {\em Let $n$ as above be the length. Then we can construct an Hermitian dual-containing BCH code ${\bf C}$ with the defining set ${\bf T}=C_1 \cup C_2 \cup \cdots \cup C_{\delta-1}$, where $\delta = \frac{q^m-1}{\mu}$. Then we have $$d({\bf C}) \geq \frac{q^m-1}{\mu}+1.$$}\\

\section{Euclidean and Hermitian self-dual cyclic codes over ${\bf F}_{2^s}$ with large minimum distances}

From Theorem 2.1, Theorem 3.2 and 3.3, we have the following construction of Euclidean and Hermitian self-dual cyclic codes with square-root-like minimum distances.\\

{\bf Theorem 4.1.} {\em Let $n=\frac{2(2^{sm}-1)}{\mu}$, where $m=3, 5, 7, \ldots$ and $\mu$ is a divisor of $2^{sm}-1$. We can construct a family of Euclidean self-dual repeated-root cyclic codes over ${\bf F}_{2^s}$ with the length $n$ and the minimum distances at least $\sqrt{\frac{2^{s-1}n}{\mu}}-\frac{2^s}{\mu}$.}\\

{\bf Proof.} From Theorem 3.2 we have a dual-containing BCH code ${\bf C} \subset {\bf F}_2^{2^{sm}-1}$ with the minimum distance at least $\frac{2^{\frac{s(m+1)}{2}}-2^s}{\mu}$. Then from the construction in Theorem 2.1 and the van Lint theorem, the code ${\bf C}_1 \subset {\bf F}_2^{2(2^{sm}-1)}$ is an Euclidean self-dual repeated-root cyclic code of the length $2n$. The minimum distance of ${\bf C}_1$ is at least $\sqrt{\frac{2^{s-1}n}{\mu}}-\frac{2^s}{\mu}$. \\

The following result follows from Theorem 4.1 directly.\\

{\bf Corollary 4.1.} {\em Let $n=2(2^{sm}-1)$, where $m=3, 5, 7, \ldots$. We can construct a family of Euclidean self-dual repeated-root cyclic codes over ${\bf F}_{2^s}$ with the length $n$ and the minimum distance at least $\sqrt{2^{s-1}n}-2^s$.}\\

For Hermitian self-dual cyclic codes, we have the following results.\\

{\bf Theorem 4.2.} {\em Let $n=\frac{2(2^{2sm}-1)}{\mu}$, where $m=3, 5, 7, \ldots$ and $\mu$ is a divisor of $2^{2sm}-1$. We can construct a family of Hermitian self-dual repeated-root cyclic codes over ${\bf F}_{2^{2s}}$ with the length $n$ and the minimum distances at least $\sqrt{\frac{n}{2\mu}}$.}\\

{\bf Corollary 4.2.} {\em Let $n=2(2^{2sm}-1)$, where $m=3, 5, 7, \ldots$ and $s=1, 2, \ldots$. We can construct a family of Hermitian self-dual repeated-root cyclic codes over ${\bf F}_{2^{2s}}$ with the length $n$ and the minimum distance at least $\sqrt{n/2}$.}\\

The automorphism groups of codes constructed in above results contain the subgroup ${\bf Z}_{n/2} \times {\bf Z}_{n/2}$.\\

\section{Conclusions}

The construction of self-dual codes over small fields such that their minimum distances are as large as possible has a long history in the coding theory. New families of Euclidean and Hermitian self-dual repeated-root cyclic codes over ${\bf F}_{2^s}$, $s \geq 2$,  with square-root-like minimum distances, were constructed in this paper. Our results improved results in two papers \cite{HTop,TangDing} published in IEEE Trans. Inf. Theory significantly. \\

\end{document}